\documentclass[]{iopart}
\usepackage{iopams}
\usepackage{setstack}
\usepackage[utf8]{inputenc}
\usepackage[T1]{fontenc}
\usepackage{lmodern}
\usepackage[english]{babel}
\usepackage[dvips]{graphicx}
\usepackage{color}

\newcommand{\NA}{N_{\rm A}}
\newcommand{\dd}{d_{220}}

\begin{document}

\title[]{Finite element analysis of surface-stress effects in the Si lattice-parameter measurement}
\author{C Sasso$^1$, D Quagliotti$^1$, E Massa$^1$, G Mana$^1$ and U Kuetgens$^2$}
\address{$^1$INRIM -- Istituto Nazionale di Ricerca Metrologica, str.\ delle Cacce 91, 10135 Torino, Italy}
\address{$^2$PTB -- Physikalisch-Technische Bundesanstalt, Bundesallee 100, 38116 Braunschweig, Germany}
\ead{d.quagliotti@inrim.it}

\begin{abstract}
A stress exists in solids surfaces, similarly to liquids, also if the underlying bulk material is stress-free. This paper investigates the surface stress effect on the measured value of the Si lattice parameter used to determine the Avogadro constant by counting Si atoms. An elastic-film model has been used to provide a surface load in a finite element analysis of the lattice strain of the x-ray interferometer crystal used to measure the lattice parameter. Eventually, an experiment is proposed to work a lattice parameter measurement out so that there is a visible effect of the surface stress.
\end{abstract}

\submitto{Metrologia}




\section{Introduction}
The Avogadro constant, $\NA$, is determined by dividing the molar volume of a silicon single-crystal highly enriched with $^{28}$Si by its unit cell volume, obtained from the measured value of the lattice parameter \cite{Andreas:2011a,Andreas:2011b,Massa:2011}. The lattice parameter is measured by means of combined x-ray and optical interferometry. Recently, the effect of surface stress was brought to our attention as a possible error source in our measurement of the lattice parameter \cite{Kren:2012}. Indeed, this issue was already taken into account but never definitely investigated  \cite{Ferroglio:2008}. An intrinsic surface-stress exists in solid, in a similar way as it occurs in liquids, which was first modelled by Gibbs \cite{Gibbs:1906} and Shuttleworth \cite{Shuttleworth:1950}, whose works were seminal in surface thermodynamics. Though it is widely reported in review papers and textbooks on surface science \cite{Cammarata:1994,Desjonqueres:1996}, their theory is still a matter of debate \cite{Makkonen:2012}. A continuous mechanics theory of solid surfaces has been developed by Gurtin and Murdoch \cite{Gurtin:1975,Gurtin:1978}, which is summarized in \cite{Wolfer:2011}. When the surface relaxes, it compresses or pulls the underlying crystal; since the resulting lattice strain depends on the crystal size and geometry, the lattice parameter of an x-ray interferometer is different from that of the Si spheres used to determine the molar volume.

After summarizing the operation of a combined x-ray and optical interferometer, this paper investigates the effect of surface stress on the interferometer operation. An elastic membrane model has been used to provide a surface load in a finite element analysis of the lattice strain. Particular emphasis was given to the operation of the $^{28}$Si interferometer used to determine $\NA$. Eventually, a measurement of the lattice parameter with a variable-thickness interferometer is proposed in order to evidence the effect of surface stress on the underlying crystal.

\begin{figure}
\centering
\includegraphics[width=65mm]{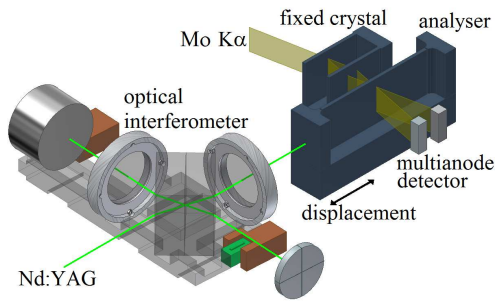}
\caption{Sketch of a combined x-ray and optical interferometer. The analyser crystal moves in a direction parallel to the laser beam, its displacement is measured by the optical interferometer, and each passing diffracting plane is counted. Both the x-ray and optical interference patterns are imaged into position-sensitive detectors to sense lattice strain and parasitic rotations of the analyser.}\label{XINT}
\end{figure}

\section{Combined x-ray and optical interferometry}
A review of the lattice parameter measurement can be found in \cite{Becker:1994}. As shown in Fig.~\ref{XINT}, an x-ray interferometer consists of three Si crystal slabs so cut that the \{220\} planes are orthogonal to the crystal surfaces. X-rays from a 17~keV~Mo~K$\alpha$ source are split by the first crystal and recombined, via two transmission crystals, by the third, called the analyser. When the analyser is moved in a direction orthogonal to the \{220\} planes, a periodic intensity-variation of the transmitted and diffracted x-rays is observed, the period being the diffracting-plane spacing, about 192~pm. The analyser embeds front and rear mirrors, so that its displacement and rotations can be measured by an optical interferometer. The measurement equation is $\dd = m \lambda/(2n)$, where $\dd \approx 192$~pm is the spacing of the \{220\} planes and $n$ is the number of x-ray fringes observed in a displacement of $m$ optical fringes having period $\lambda/2 \approx 266$~nm. To ensure the interferometer calibration, the laser source operates in single-mode and its frequency is stabilized against that of a transition of the $^{127}$I$_2$ molecule. To eliminate the adverse influence of the refractive index of air and to achieve millikelvin temperature uniformity and stability, the experiment is carried out in a thermovacuum chamber. Continuous developments led the measurement sensitivity and accuracy to approach $10^{-9}\dd$; the most recent results are given in \cite{Massa:2011}.

\section{Surface stress}
Similarly to liquid surfaces, solid surfaces exhibit a stress also if the bulk material is stress-free. This is the macroscopic consequence of atomic affects: roughly speaking, the atoms on the surface lose half of the bonds they would have in the bulk; therefore, crystal surfaces are disturbances in the bonding of atoms.

Surface stress was first generalized to solids by Gibbs and Shuttleworth \cite{Gibbs:1906,Shuttleworth:1950} as the amount of reversible work per unit area needed to stretch a pre-existing surface. Since a solid surface may change not only by exchange of matter with the bulk, but also by elastic deformation, the concept of surface stress must be derived from the molecular dynamics of the surface \cite{Miyamoto:1994,Hara:2005}.

Continuum mechanics provides the framework to define the surface properties and the relationships among them. The elasticity theory of solid surfaces to which we refer was developed by Gurtin and Murdoch \cite{Gurtin:1975,Gurtin:1978}. A summary of the linearised theory -- which is valid when both the surface and bulk strains are small -- is given in \cite{Wolfer:2011}. The key concept is that, for a surface attached to the underlying solid, a stress-free reference does not exist; since the bonds between the surface atoms and between the bulk atoms are different, the surface of a stress-free crystal has residual strain, $\epsilon_0^{\alpha\beta}$, and stress, $\sigma_0^{\alpha\beta}$.

If the surface tends to shrink (expand) with respect to the bulk, the surface stress is positive (negative) and it is said to be tensile (compressive). Since a fully relaxed surface has no normal stress, surface strain and stress are second rank tensors, which are characterized by two principal orthogonal-components. If the surface has more than a 2-fold symmetry, the surface stress and strain tensors are isotropic. In this case, $\epsilon_0^{\alpha\beta}= \epsilon_0\delta_{\alpha\beta}$ and the free surface-energy per unit area is \cite{Wolfer:2011}
\begin{equation}
 \gamma = \gamma_0 +2(\mu+\lambda)\epsilon_0\epsilon_{\alpha\alpha} + \mu\epsilon_{\alpha\beta}\epsilon_{\alpha\beta} +
 \lambda\epsilon_{\alpha\alpha}\epsilon_{\beta\beta}/2 ,
\end{equation}\label{free}
where $\epsilon_{\alpha\beta}$ are the strain component in the plane tangent to the surface, $\gamma_0$ is the surface energy for a stress-free bulk, $\mu$ and $\lambda$ are elastic constants, and the summation over the repeated indices -- $\alpha, \beta = 1,2$ -- is implied. The definition of surface stress is $\sigma_{\alpha\beta}=\partial\gamma / \partial\epsilon_{\alpha\beta}$; hence, by taking the derivatives,
\begin{equation}\label{stress}
 \sigma_{\alpha\beta} = \sigma_0\delta_{\alpha\beta} + 2\mu\epsilon_{\alpha\beta} +
 \lambda(\epsilon_{\alpha\alpha}+\epsilon_{\beta\beta})\delta_{\alpha\beta}/2 ,
\end{equation}
where $\sigma_0=2(\mu+\lambda)\epsilon_0$ is the intrinsic surface stress.

\begin{table}
\caption{\label{s-stress} Calculated values of the principal components, $\sigma_\|$ and $\sigma_\bot$, and mean value, $\sigma_{001}$, of the surface stress of the symmetrically dimerized Si(001) $(2\times 1)$ surface. The $\|$ and $\bot$ directions are parallel and orthogonal to the dimer rows; positive and negative values indicate tensile and compressive stresses.}
\begin{indented}
\item[]\begin{tabular}{@{}rrrll}
\br
 $\sigma_\|$/(N/m) &$\sigma_\bot$/(N/m) &$\sigma_{001}$/(N/m) &year &reference\\
\mr
$0.58$  &$-1.28$ &$-0.35$   &1988 &\cite{Alerhand:1988} \\ 
$0.75$  &$-2.11$ &$-0.68$   &1989 &\cite{Payne:1989}    \\ 
$1.70$  &$-0.96$ &$ 0.37$   &1990 &\cite{Meade:1990}    \\
$1.18$  &$-0.05$ &$ 0.57$   &1992 &\cite{Poon:1992}     \\ 
$2.38$  &$-0.86$ &$ 0.76$   &1993 &\cite{Garcia:1993}   \\ 
$0.75$  &$-1.18$ &$-0.22$   &1994 &\cite{Miyamoto:1994} \\ 
$0.73$  &$-1.06$ &$-0.17$   &2005 &\cite{Hara:2005}     \\ 
$0.40$  &$-1.34$ &$-0.47$   &2005 &\cite{Hara:2005}     \\ 
$0.13$  &$-1.12$ &$-0.50$   &2008 &\cite{Delph:2008}    \\ 
\br
\end{tabular}
\end{indented}
\end{table}

The determination of the surface stress in solids is a challenge; the measurement of the lattice parameter in small particles has been the main method \cite{Wolfer:2011}. Given the experimental difficulties and since the results are affected by the sample preparation and adsorbed impurities, the values obtained from {\it ab-initio} and molecular dynamics calculations seem more reliable. The results for the symmetric $(2\times 1)$ reconstruction of the Si(001) surface are summarized in table~\ref{s-stress}; the reconstructed surface is formed by domains of anisotropic stress, but, for symmetry reasons, the mean stress is isotropic. The scatter of data suggests a reassuring small surface stress.

Because of the anisotropic nature of the crystal, surface stress depends on the surface orientation. A measurement of the polar dependence of the relative surface-stress at high temperature (1373 K) is given in \cite{Metois:2005}, which allows the surface stress of any surface to be determined from that of the (001) reference surface.  In the case of our x-ray interferometer, the $(\overline{1}10)$ surfaces are those of main interest; because of their 2-fold symmetry, two values are necessary to represent the surface stress of the two perpendicular $[001]$ and $[110]$ components. As reported in \cite{Metois:2005}, at 1373 K, the sought principal components are $\sigma_{\overline{1}10}^{001} \approx 0.5\sigma_{001}$ and $\sigma_{\overline{1}10}^{110}\approx 0.3 \sigma_{001}$. Indicative $\sigma_{001}$ values are given in table \ref{s-stress}.

The crystal's surfaces are normally covered by a few nanometre thick oxide-layer, which can further stress the surface. The stress changes of the Si(001) $(2\times 1)$ surface during plasma oxidation were measured by using a thin silicon cantilever; the bending caused by the oxidation of one face was optically measured and related to the surfaces stress difference by elasticity theory \cite{Itakura:2000}. Initially, the oxidation induces a compressive stress, followed by a tensile stress; eventually, the stress variation turns to compressive for oxide thickness greater than about 1.5~nm. The initial stress changes were explained in terms of the electron diffusion into the sample and the shrinkage caused by the oxygen-bridged dimer structure. The final compressive variation, $\Delta \sigma_{001} \approx -0.1$~N/m, was explained by the oxidation of Si atoms at deep sites; since the oxygen expands the Si--Si bond, oxidation is expected to cause a compressive stress.

The surface relaxes the intrinsic stress by compressing or pulling the underlying crystal. For instance, the (isotropic) lattice strain of a sphere of radius $R$ is
\begin{equation}\label{sphere}
 \epsilon = -\frac{2\sigma_0}{4(\mu+\lambda)+3KR} \approx -\frac{2\sigma_0}{3KR},
\end{equation}
where $K$ is the bulk modulus and the Young-Laplace approximation holds when the elastic constants can be neglected when compared to the $KR$ product. By choosing $\sigma_0=1$~N/m, we can use this formula -- where $2R=93$~mm and $K=98$~GPa -- to estimate the lattice strain of the spheres used to determine $\NA$. The result is $\epsilon=-0.07$~nm/m, which is irrelevant for the $\NA$ determination. However, if $2R=1.2$~mm -- which is a value equal to the thickness of the crystal used to measure the lattice parameter -- we obtain $\epsilon=-6$~nm/m, which is significant with respect to  the uncertainty of the  lattice-parameter measurement.

\section{Finite element analysis}
\begin{figure}
\includegraphics[width=65mm]{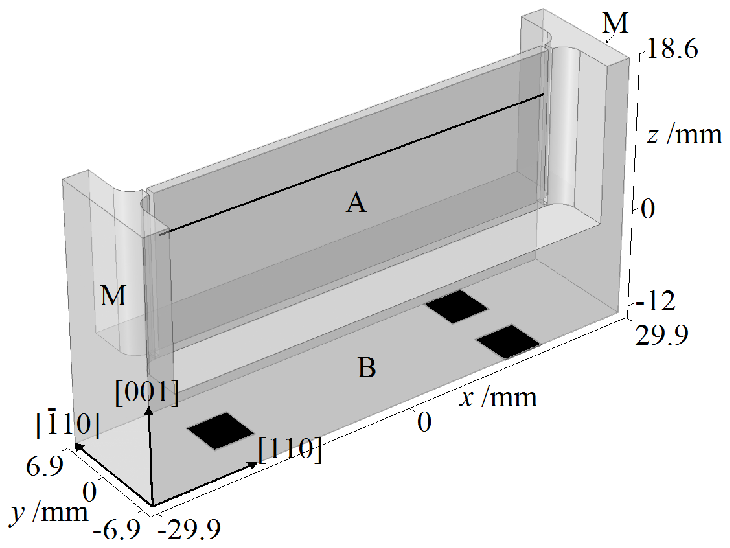}
\includegraphics[width=65mm]{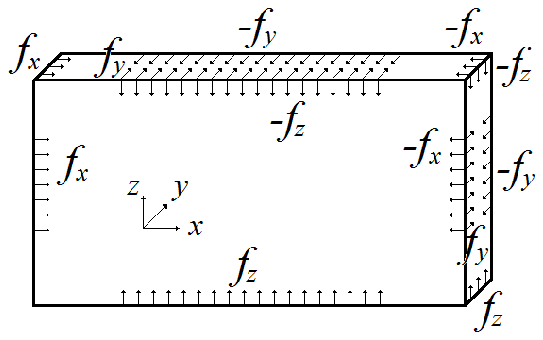}
\caption{Left: sketch of the analyser crystal of the $^{28}$Si interferometer. M: end mirrors, A: analyser lamella, B: base. The $(5 \times 5 \times 0.1)$~mm$^3$ reliefs that support the crystal are coloured in black. The lattice parameter was measured along the indicated line, at $z=14$~mm. Right: surface-stress modelling by shear forces.}\label{analyser}
\end{figure}

Figure~\ref{analyser} (left) shows the analyser crystal of the $^{28}$Si x-ray interferometer used for the $\NA$ measurement and its crystallographic orientation. The end mirrors are parallel to the diffracting planes; to avoid stress propagation, they are separated by vertical cuts. The lamella thickness, 1.2~mm, was chosen by a trade off between the increased lattice strain (due to the surface stress) of a thin lamella and the increased x-ray absorption of a thick lamella.

To estimate the lattice parameter change caused by the surface stress, we used a commercial finite element analysis software \cite{comsol}. As shown in Fig.\ \ref{analyser}, the (220) planes are orthogonal to the lamella faces which are parallel to $(\overline{1}10)$ planes. Owing the large uncertainty of the residual stress values, though the $(\overline{1}10)$ surfaces are anisotropic, the analyser surfaces were modelled as stretched by isotropic membranes attached to the underlying crystal lattice. Since we are only concerned with infinitesimal strains, though the surface stress $\sigma_{\alpha\beta}$ depends on the strain according to (\ref{stress}), we used an isotropic $\sigma_0=1$~N/m value. This surface stress value allows the analysis results to be easily scaled if a different value has to be considered and, hopefully, it is larger than the actual stress value. Hence, according the Young-Laplace equation, the surface applies the inward pressure $P=2\kappa\sigma_0$, where $\kappa$ is the mean curvature. As schematically shown in Fig.~\ref{analyser} (right), when considering flat surfaces, the surface stress was modelled by shear forces per unit length applied orthogonally to the surface edges.

As strains are infinitesimal, we uses a linear model; the mesh, of about $10^6$ tetrahedral anisotropic elements, was the result of successive refinements. The not null elements of the stiffness matrix are given table~\ref{S:matrix} in both the [100], [010], [001] and [110], [$\overline{1}$10], [001] crystal axes. As regards the model parameters, the Si density and gravitational acceleration are 2330~kg/m$^3$ and 9.81~m/s$^2$, respectively.

\begin{table}[b]
\caption{\label{S:matrix} Not null elements of the stiffness matrix for a silicon crystal. The measurement unit is GPa.}
\begin{indented}
\item[]\begin{tabular}{@{}ccccccccc}
\br
\multicolumn{9}{c}{[100], [010], [001] axes} \\
\multicolumn{3}{c}{$c_{11}, c_{22}, c_{33}$} & \multicolumn{3}{c}{$c_{12}, c_{13}, c_{23}$} & \multicolumn{3}{c}{$c_{44}, c_{55}, c_{66}$}\\
\mr
\multicolumn{3}{c}{165.7} & \multicolumn{3}{c}{63.9} & \multicolumn{3}{c}{79.6}\\
\br
\multicolumn{9}{c}{[110], [$\overline{1}$10], [001] axes} \\
\multicolumn{2}{c}{$c^,_{11},c^,_{22}$} & $c^,_{12}$ & \multicolumn{2}{c}{$c^,_{13},c^,_{23}$} & $c^,_{33}$ & \multicolumn{2}{c}{$c^,_{44},c^,_{55}$} & $c^,_{66}$\\
\mr
\multicolumn{2}{c}{194.4} & 35.2 & \multicolumn{2}{c}{63.9} & 165.7 & \multicolumn{2}{c}{79.6} & 50.9\\
\br
\end{tabular}
\end{indented}
\end{table}

The analyser rests on a silicon plate; it stands on three $(5 \times 5 \times 0.1)$~mm$^3$ reliefs carved out of the crystal base by chemical etching. In order to avoid stresses, it is held in position by a thin layer of high-viscosity silicone oil. Since the oil applies only viscous forces, the contact with the support has been simulated by setting to zero the displacement of the contact nodes.

\subsection{Results}
\begin{figure}
\includegraphics[width=65mm]{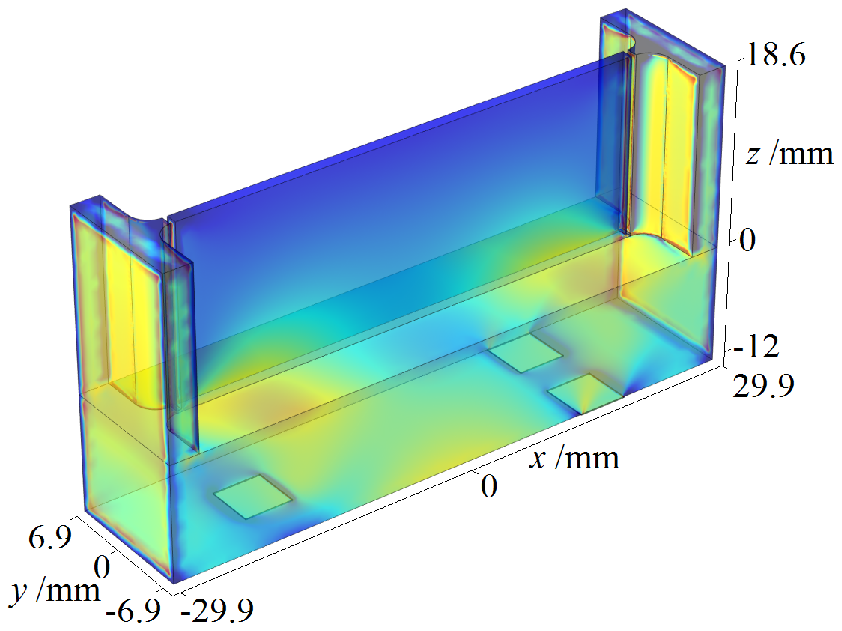}
\includegraphics[width=65mm]{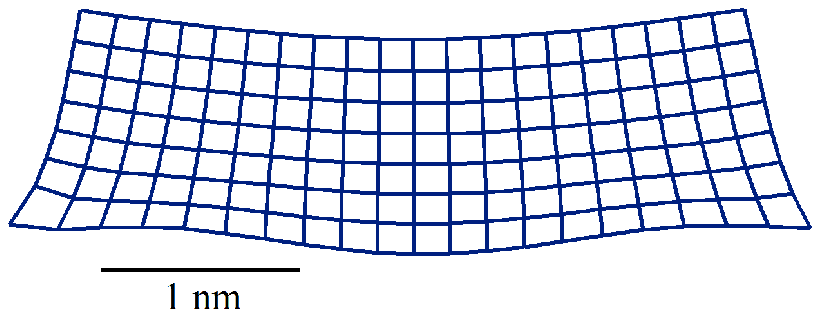}
\caption{Left: $\epsilon_{xx}$ component of the diffracting-plane strain for the optimal location of the supports, where blue is $-10$~nm/m and red is $10$~nm/m. Right: qualitative magnified image of the diffracting planes. A realignment of the analyser has been simulated by subtracting the average displacement and tilt. The displacement magnification factors is shown by the 1~nm bar.}\label{strain:1}
\end{figure}

We calculated the strain $\epsilon_{xx}=\partial_x u_x$, where $u_x$ is the horizontal component of the displacement vector. Both the effects of self-weight and surface strain have been considered; the supports were so chosen as to minimize bending or sagging \cite{Ferroglio:2011}. The three dimensional map of the strain distribution in the analyser and a magnified image of the strained diffracting planes are shown in Fig.~\ref{strain:1}.

\begin{figure}[b]
\includegraphics[width=65mm]{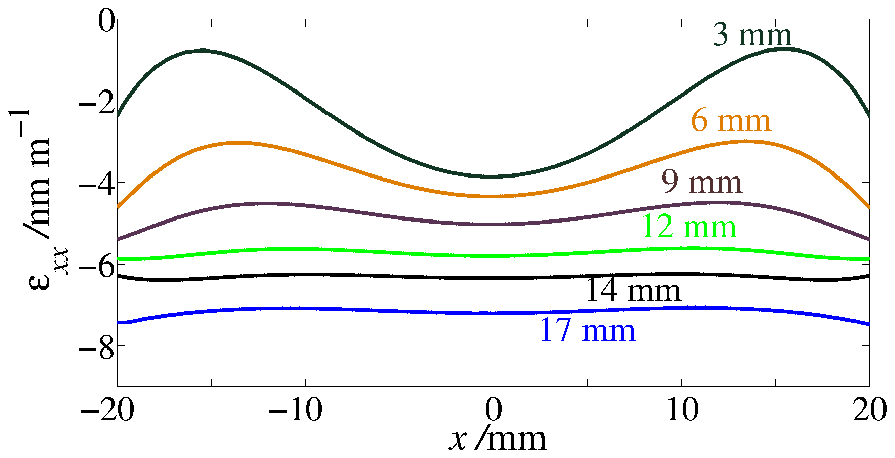}
\includegraphics[width=65mm]{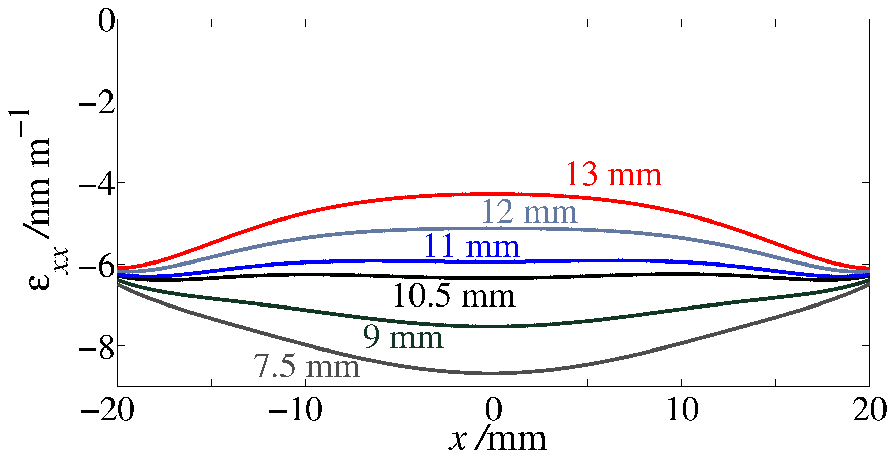}
\caption{Left: horizontal $\dd$ variations in $y = 0$ plane, at different distances from the analyser bottom-surface. Right: Relative $\dd$ variations along the $z=14$~mm measurement baseline (figure~\ref{analyser}). Different support-to-mirror distances have been considered, as indicated in the graph.}\label{strain:2}
\end{figure}

The $\epsilon_{xx}$ strain is trivially related to the lattice parameter by $\Delta \dd/\dd = \epsilon_{xx}$. Hence, the relative variations of the $\dd$ values along measurement lines at different heights are identical to the $\epsilon_{xx}$ values calculated along the same lines; they are shown in Fig.\ \ref{strain:2} (left). The figure indicates that, in the measurement horizontal strip a couple of millimetre wide at a $z=14$~mm, the lattice is uniformly strained by about $-6$~nm/m, as estimated by the previous ``spherical-cow'' model. In addition to the horizontal strain, there is a vertical strain gradient of about $3$~nm~m$^{-1}$~cm$^{-1}$, the bottom part being less strained, which could be the subject matter of future experimental verifications.

Figure~\ref{strain:2} (right) updates the corresponding figure~6 of \cite{Ferroglio:2011}, which was used to assess the sole effect of self weigh in the lattice parameter measurement. It shows that, if the analyser surfaces are loaded by a tensile stress of 1~N/m, the $\dd$ value is about $6 \times 10^{-9}\dd$ smaller than its value in an unstrained crystal. The figure also shows that the lattice parameter profile given in \cite{Massa:2011} cannot deliver clues about the surface contribution to the measured $\dd$ value because the only effect of the surface stress is to translate all the curves downwards by $6 \times 10^{-9}\dd$.

Since the uncertainty associated to the measured $\dd$ value is $3.2\times 10^{-9}\dd$, the lattice strain induced by a 1 N/m stress impairs the measurement accuracy. The results indicate also that we can cope effectively the surface stress by a numerical determination of its effect. However, owing the large uncertainty of the sign and value of the surface stress, we don't yet propose to reconsider the corrections and error-budget contributions given in \cite{Massa:2011}. Further investigations are under way to assess and to quantify the effect of surface stress, if any. In the next section we describe an analyser design to work a lattice parameter measurement out so that there is a measurable effect of the surface stress.

\begin{figure}
\centering
\includegraphics[width=65mm]{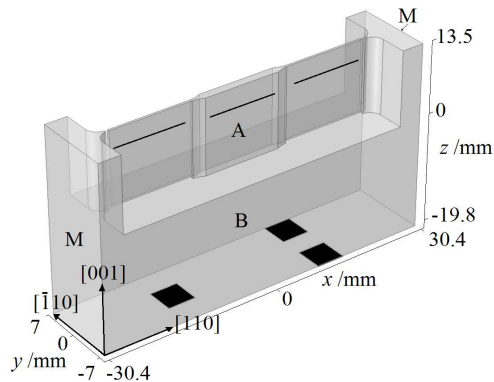}
\caption{Sketch of the variable-thickness analyser; vertical and horizontal cuts separate the end, thin, flag-like lamellae from the mirrors and base. The expected $\dd$ variations are calculated along the indicated lines, at $z=9$~mm.} \label{new-XINT}
\end{figure}

\section{Variable thickness interferometer}
A way to evidence if the surface stress strains the analyser crystal up to an extent which is significant to the lattice parameter measurement is to compare the results of measurements carried out in crystals having a large thickness difference. Since the induced $\dd$ changes are small, to avoid they are masked by lattice imperfections, measurements must be carried out in the same crystal. Consequently, as shown in Fig.~\ref{new-XINT}, we designed an x-ray interferometer having a variable analyser thickness. The minimum 0.4~mm thickness is set by machining capabilities; the maximum 1.5~mm thickness is set by x-ray absorption.

\begin{figure}
\includegraphics[width=65mm]{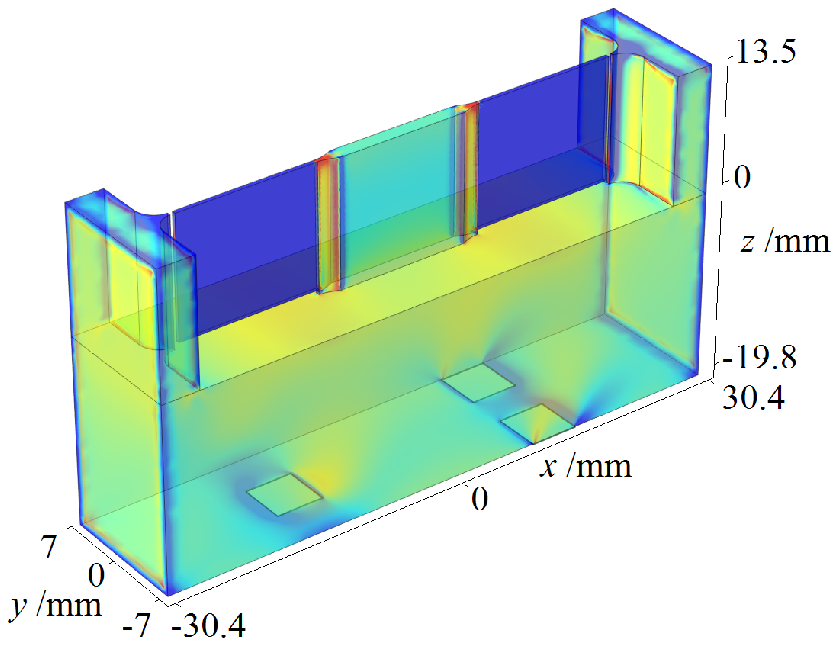}
\includegraphics[width=65mm]{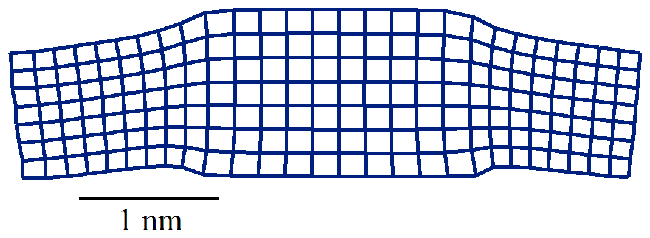}
\caption{Left: $\epsilon_{xx}$ component of the diffracting-plane strain for the variable-thickness analyser, where blue is $-10$~nm/m and red is $10$~nm/m. Right: qualitative magnified image of the diffracting planes. A realignment of the analyser has been simulated by subtracting the average displacement and tilt. The displacement magnification factors is shown by the 1~nm bar.}\label{new-strain:1}
\end{figure}

\begin{figure}[b]
\includegraphics[width=65mm]{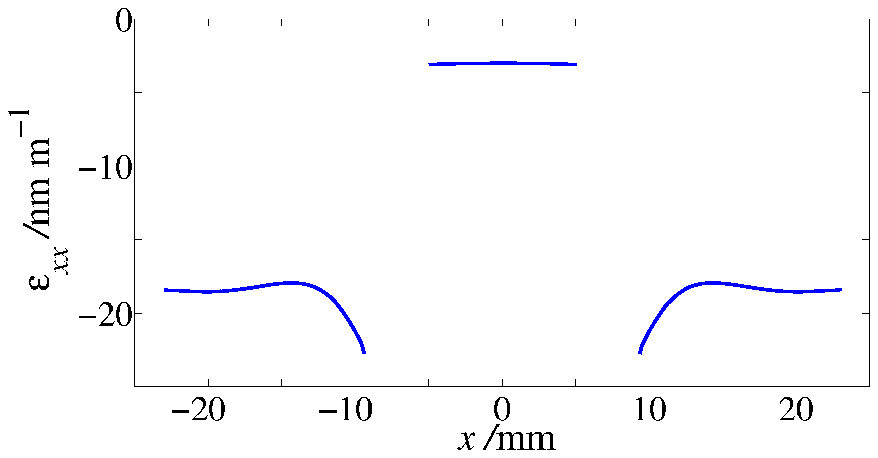}
\includegraphics[width=65mm]{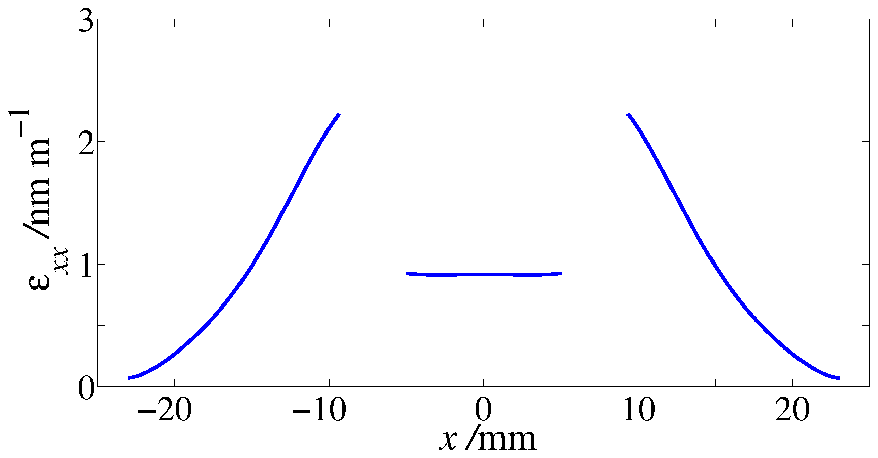}
\caption{Horizontal $\dd$ variations in $y = 0$ plane, along the $z=9$~mm measurement lines (figure~\ref{new-XINT}). The supports are optimally located so as to minimize the self-weight bending. Left: Both the self-weight and surface stress are considered. Right: Only the self-weight is considered.}\label{new-strain:2}
\end{figure}

After the optimization of the finite element analysis and the location of the supports so as to minimize the self-weight bending, we obtained the strain distribution shown in Fig.~\ref{new-strain:1} (left); a magnified image of the strained \{220\} planes is shown in Fig.~\ref{new-strain:1} (right). The expected $\dd$ profile along the horizontal lines at $z=9$~mm height is shown in Fig.~\ref{new-strain:2} (left). As shown in Fig.\ \ref{new-strain:2} (right), the surface stress and self-weight have opposite effects.

\begin{figure}
\centering
\includegraphics[width=65mm]{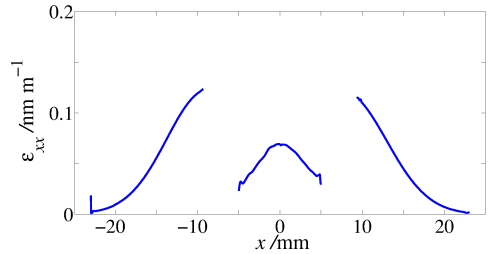}
\caption{Difference between the horizontal $\dd$ variations in $y = 0$ plane, along the $z=9$~mm measurement lines, when the analyser supports are displaced (in $x-y$ plane) from their optimal positions up to $\pm1$~cm.} \label{rejection}
\end{figure}

The surface stress will be estimated from the $\dd$ gap between the thick and thin analyser parts. If the surface is loaded by a tensile stress of 1~N/m, $\dd$ decreases by about $15 \times 10^{-9}\dd$ from the central (thick) to the outer (thin) parts. This variation is large enough to be detected. With a high crystal perfection, the present detection limit of $3 \times 10^{-9}\dd$ corresponds to a minimum detectable stress of 0.2~N/m. To ensure that self-weight bending does not change in a significant way the $\dd$ gap between the thick and thin analyser parts, the height of the analyser base was increased to 19.8~mm. Figure~\ref{rejection} shows that, when the analyser supports are randomly displaced (in $x-y$ plane) from their optimal positions up to $\pm1$~cm, the $\dd$ gap between the thick and thin analyser-parts changes less than what can be experimentally detected.

\section{Conclusions}
The influence of the intrinsic surface-stress on the operation of the x-ray interferometer utilized to measure the lattice parameter of the $^{28}$Si crystal used to determine the Avogadro constant has been investigated by finite element analysis. The crystal surfaces, modelled as membranes having a tensile stress of 1~N/m, relax by compressing the underlying crystal. Though the induced strain is greater than the uncertainty associated to the measured value of the lattice parameter, since the stress magnitude is highly uncertain and the chosen 1~N/m value could be pessimistic, we do not propose a correction of the published value. To verify if the surface stress contributes or does not contribute to the lattice parameter measurement, a variable-thickness design of the x-ray interferometer has been investigated to induce detectable strain variations in the same interferometer crystal. The realization of such an interferometer is under way in collaboration at the Physikalisch Technische Bundesanstalt in collaboration with the Leibniz-Institut f\"ur Oberfl\"achenmodifizierung.

\ack
We thank Petr Kren of the Czech Metrology Institute for having brought to our attention this criticality of the lattice parameter measurement. This work was jointly funded by the European Metrology Research Programme (EMRP) participating countries within the European Association of National Metrology Institutes (EURAMET) and the European Union.

\end{document}